\documentclass[jcp,twocolumn,superscriptaddress,floatfix,graphicx,showpacs]{revtex4-1}
\usepackage{ifpdf}
 \newif\ifpdf
\ifx\pdfoutput\undefined
   \pdffalse
\else
   \pdfoutput=1
   \pdftrue
\fi
\ifpdf
   \usepackage{graphicx}
   \usepackage{epstopdf}
   \usepackage{color}
\else
   \usepackage{graphicx}
\fi

\usepackage{srctex}
\usepackage{hyperref}
\usepackage{natbib}
\usepackage{amsmath,amssymb}

\begin{document}

\title{The uncertainty of glass transition temperature in molecular dynamics simulations and numerical algorithm for its unique determination}


\author{N.M. Chtchelkatchev}
\affiliation{Institute for High Pressure Physics, Russian Academy of Sciences, 142190 Troitsk, Russia}

\author{R.E. Ryltsev}
\affiliation{Institute of Metallurgy, Ural Branch of Russian Academy of Sciences, 620016, 101 Amundsena str., Ekaterinburg, Russia}
\affiliation{Ural Federal University, 620002, 19 Mira str., Ekaterinburg, Russia}
\affiliation{Institute for High Pressure Physics, Russian Academy of Sciences, 142190 Troitsk, Russia}

\author{V. Ankudinov}
\affiliation{Institute for High Pressure Physics, Russian Academy of Sciences, 142190 Troitsk, Russia}

\author{V.N. Ryzhov}
\affiliation{Institute for High Pressure Physics, Russian Academy of Sciences, 142190 Troitsk, Russia}

\author{M. Apel}
\affiliation{Access e.V. – Materials and Processes
An-Institut der RWTH Aachen, 52074 Aachen, Germany}

\author{P.K. Galenko}
\affiliation{Faculty of Physics and Astronomy, Otto Schott Institute of Materials Research, Friedrich-Schiller-Universität-Jena, 07743 Jena, Germany}

\begin{abstract}
When the cooling rate $v$ is smaller than a certain material-dependent threshold, the glass transition temperature $T_g$ becomes to a certain degree the ``material parameter'' being nearly independent on the cooling rate. The common method to determine $T_g$ is to extrapolate viscosity $\nu$ of the liquid state at temperatures not far above the freezing conditions to lower temperatures where liquid freezes and viscosity is hardly measurable. It is generally accepted that the glass transition occurs when viscosity drops by $13\leq n\leq17$ orders of magnitude. The accuracy of $T_g$ depends on the extrapolation quality. We propose here an algorithm for a unique determining of $T_g$. The idea is to unambiguously extrapolate $\nu(T)$ to low temperatures without relying upon a specific model. It can be done using the numerical analytical continuation of $\nu(T)$-function from above $T_g$ where it is measurable, to $T\gtrsim T_g$. For numerical analytical continuation, we use the Pade approximant method.
\end{abstract}

\maketitle

\section{Introduction}

Understanding processes of glass formation and jamming, and predicting the corresponding conditions, are among the main tasks in chemical physics. The question ``why some liquids form a glass easily but others do not'' has been thoroughly investigated \cite{Angell1996JPC,Cheng2011ProgMateSci,Ryltsev2013PRL,Royall2015PhysRep}. However, despite the decades of researches, there is still an unsolved problem related to uncertainty in the definition of glass transition temperature $T_g$~\cite{Stillinger2013ARCMP}.

Due to the fact that glass formation is dynamical non-equilibrium transition, the glass transition temperature $T_g$ depends strongly on the cooling procedure~\cite{Fokin2005JNCS}. In particular, there is a semiempirical relationship for cooling rate dependence of $T_g$:
\begin{gather}\label{eqTgv}
T_g(v)=T_g^{(0)}+\Delta T_g(v), \quad \Delta T_g(v)=B/\log(v_0/v),
\end{gather}
Here $v$ is the cooling rate, while $v_0$ and $B$ are material parameters. This expression can satisfactory fit experimental data in a variety of glassformers when $v$ is varied over 3 decades.

We consider here the ``universal'' limit when the cooling rate $v\ll v_0$. Then the glass transition temperature becomes nearly independent on the cooling rate, $T_g(v)=T_g^{(0)}$, and so $T_g$ becomes to a certain degree the ``material parameter''.

The common method for estimating the glass transition temperature is to determine the kinematic viscosity $\nu$  in the liquid state at temperatures not far above the freezing conditions. Usually this is temperature range of the order of 100~K. Then the temperature dependencies of $\nu(T)$ is usually approximated by either the Arrhenius (strong glasses) or super-Arrhenius (fragile glasses) equations.  The most widely accepted approximations for super-Arrhenius behaviour are the Vogel-Fulcher-Tamman formula (VF)~\cite{Martinez-Garcia2014GvnoNature}, where
\begin{gather}\label{eqVF}
\nu\approx\nu_0^{\rm(VF)}\exp(-A/(T-T_0^{\rm(VF)})),
\end{gather}
and the power law from mode-coupling theory (MCT):
\begin{gather}\label{eqMCT}
\nu\approx\nu_0^{\rm (MCT)}(T-T_0^{\rm (MCT)})^{-\gamma},
\end{gather}
Here $\gamma>0$, $A$, $\nu_0^{\rm(VF)}$, $\nu_0^{\rm (MCT)}$, $T_0^{\rm(VF)}$ and $T_0^{\rm (MCT)}$ are usually obtained by fitting these expressions with experimental data or results of computer simulations.
For metallic glasses, the modification of VF-relation is often used~\cite{Kawamura2001MSEa}:
\begin{gather}\label{eqVF}
\nu\approx\nu_0\exp\left(-\frac A{T-T_0+\sqrt{(T-T_0)^2+GT}}\right),
\end{gather}
where $\nu_0$, $A,G$ and $T_0$  are fitting parameters. There are many other fitting equations.

Having the fitting parameters, one can estimate the glass transition temperature by extrapolating the fitted $\nu(T)$ to low temperatures above the singularity temperature $T_c$ where $\nu(T)$ diverges.

It is generally accepted that the glass transition occurs when $\nu_0/\nu=10^n$, where $13\leq n\leq17$ depending on a subjective choice. Uncertainty is built in this definition! The glass transition temperature depends on the choice of $n$ and the fitting equation. Generally, the dependence of $T_g$ from $n$ is quite weak while the choice of the extrapolating equation may sometimes give even $100\%$ uncertainty.

We propose here an algorithm for a unique determining of $T_g$. The idea is to unambiguously extrapolate $\nu(T)$ to a low-temperature region without relying upon a specific model. It can be done using the numerical analytical continuation of $\nu(T)$-function from above $T_g$ where it is measurable, to $T\gtrsim T_g$. For numerical analytical continuation, we use the Pade approximant method~\cite{baker1996pade}.

Application of Pade approximation for numerical analytical continuation in physical problems has long history of success. This method, for example, has been successfully applied for extrapolations of the bulk-magnetization data to yield the Curie point; a good agreement is obtained if one assumes that the magnetization varies near the Curie temperature~\cite{Craig1965PRB,Chakraborty1976JPhys}. Near the Curi point $T_c$ of a dilute random substitutional ferromagnetic alloy, the magnetization varies as $(T_c-T)^\alpha$ with $\alpha$ between $1/4$ and $1/3$. Such behavior suggests the possibility  to describe the region of the Curie temperature by Pade-approximant calculations based upon Ising and Heisenberg models~\cite{Craig1965PRB,Baker2012book}. Pade approximation procedure has shown great success in cluster expansion methods~\cite{Baker2012book} and Dynamical Mean Field Theory (DMFT)~\cite{Anisimov2009JPCM}. It is one of the accepted methods for numerical analytical continuation (along with the maximum entropy method~\cite{Kraberger2017PRB}) of Green functions from a discrete set of (imaginary) Matsubara frequencies to continuum of real frequencies~\cite{Georges1996RMP}. A completely different application of Pade approximants refers to an accurate description of melting curves at high pressures~\cite{Kechin2001PRB}. Pade-based Kechin equations allowed to quantify all melting curves, rising, falling, and flattening, as well as curves with a maximum, on a universal basis.

Below we describe our method and give few illustrative examples.

\begin{figure}[t]
  \centering
  \includegraphics[width=\columnwidth]{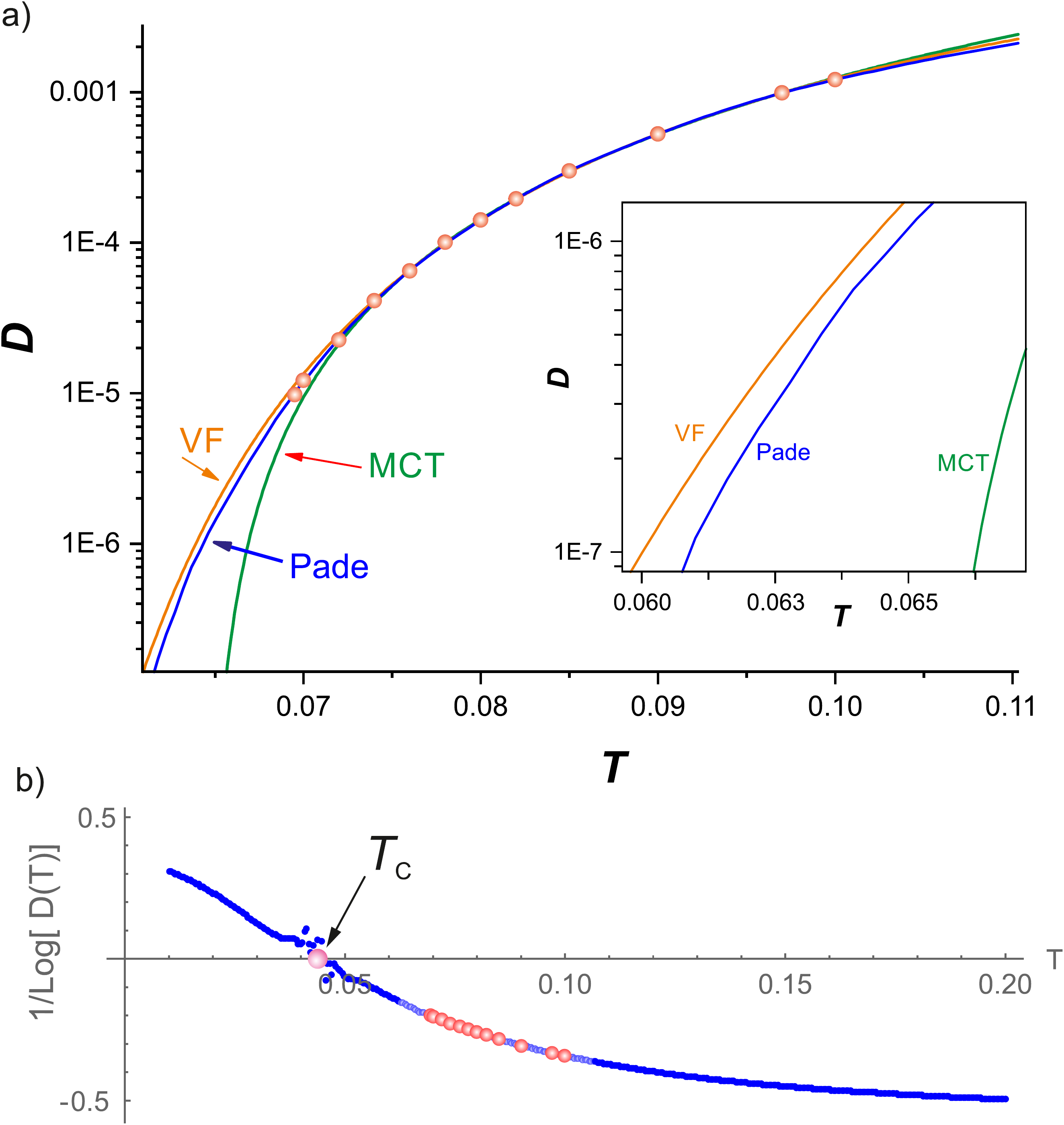}
  \caption{(Color online) Temperature dependence of diffusion coefficient  $D$ of RSS system at $\sigma=1.35$ and $\rho=0.6$. Points represent results of MD simulations. Solid lines show different approximations.  RSS is the superfragile glassformer, so VF approximation is accurate. a) MCT, PADE and VF approximations.  b) Critical temperature $T_c$ extracted from Pade approximant. }\label{Pade_vs_MCT}
\end{figure}
\begin{figure}[t]
  \centering
  \includegraphics[width=\columnwidth]{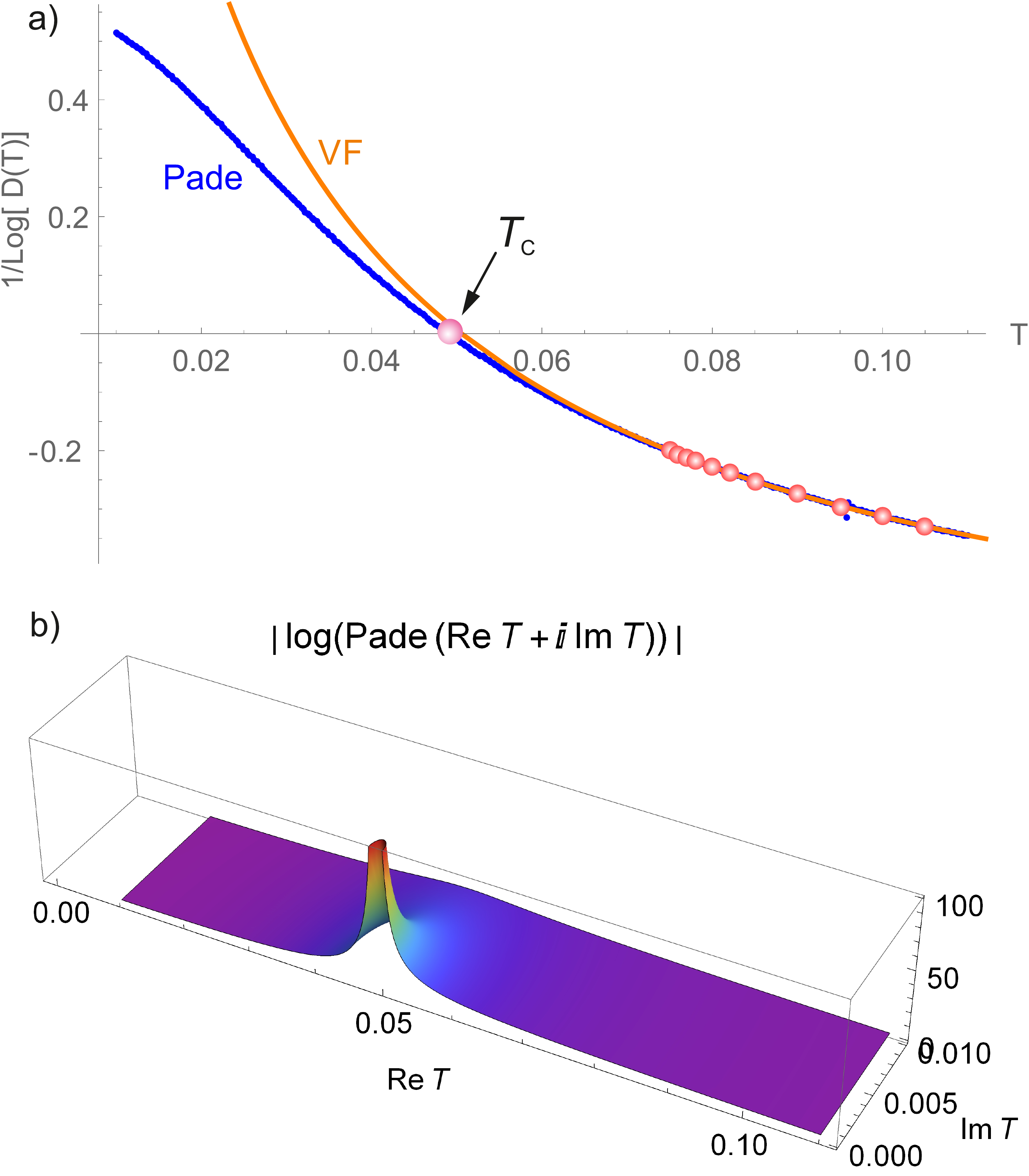}
  \caption{(Color online) Approximation and extrapolation of $D(T)$ of RSS system for $\rho=0.57$. Data obtained from MD simulations~\cite{Ryltsev2013PRL} is more accurate here than in Fig.~\ref{Pade_vs_MCT}.  a) PADE and VF approximations of  $D(T)$. b) Pade approximant extrapolated into complex plane of temperature. The pole-singularity corresponds to $T_c$.}\label{Pade_vs_MCT2}
\end{figure}
\begin{figure}[t]
  \centering
  \includegraphics[width=\columnwidth]{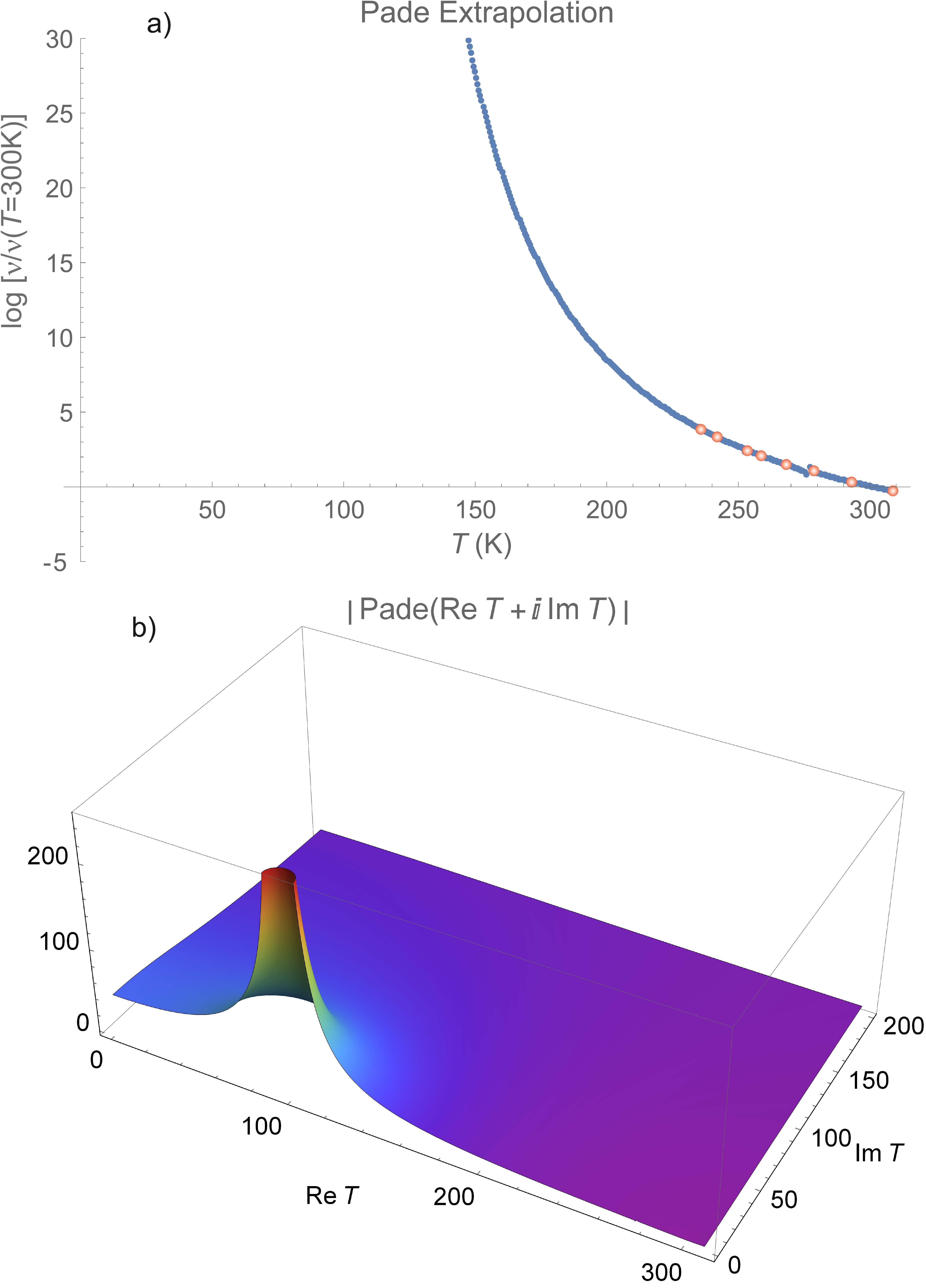}
  \caption{(Color online) a) Viscosity of glycerol in log scale. Orange spheres show experimental data and blue curve is Pade approximant. b) Pade approximant extrapolated into complex plane of temperature (temperature given in Kelvin units). The pole-singularity corresponds to $T_c$. }\label{Pade_GL}
\end{figure}
\begin{figure}[t]
  \centering
  \includegraphics[width=\columnwidth]{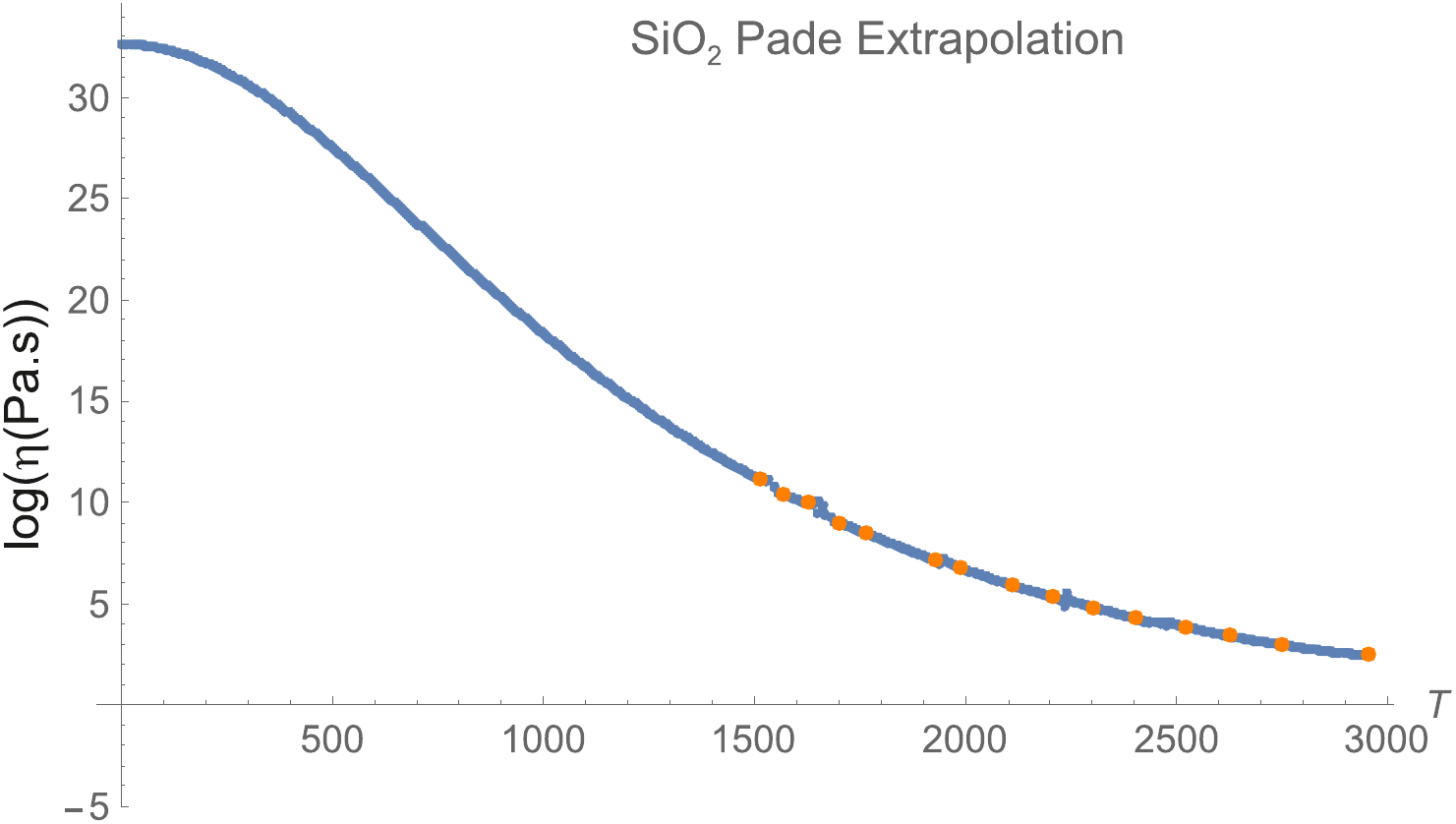}
  \caption{(Color online) Pade extrapolation of strong $\mathrm{SiO}_2$ glass (blue curve). Orange dots show experimental data.  }\label{figSiO2}
\end{figure}
\begin{figure}[t]
  \centering
  \includegraphics[width=\columnwidth]{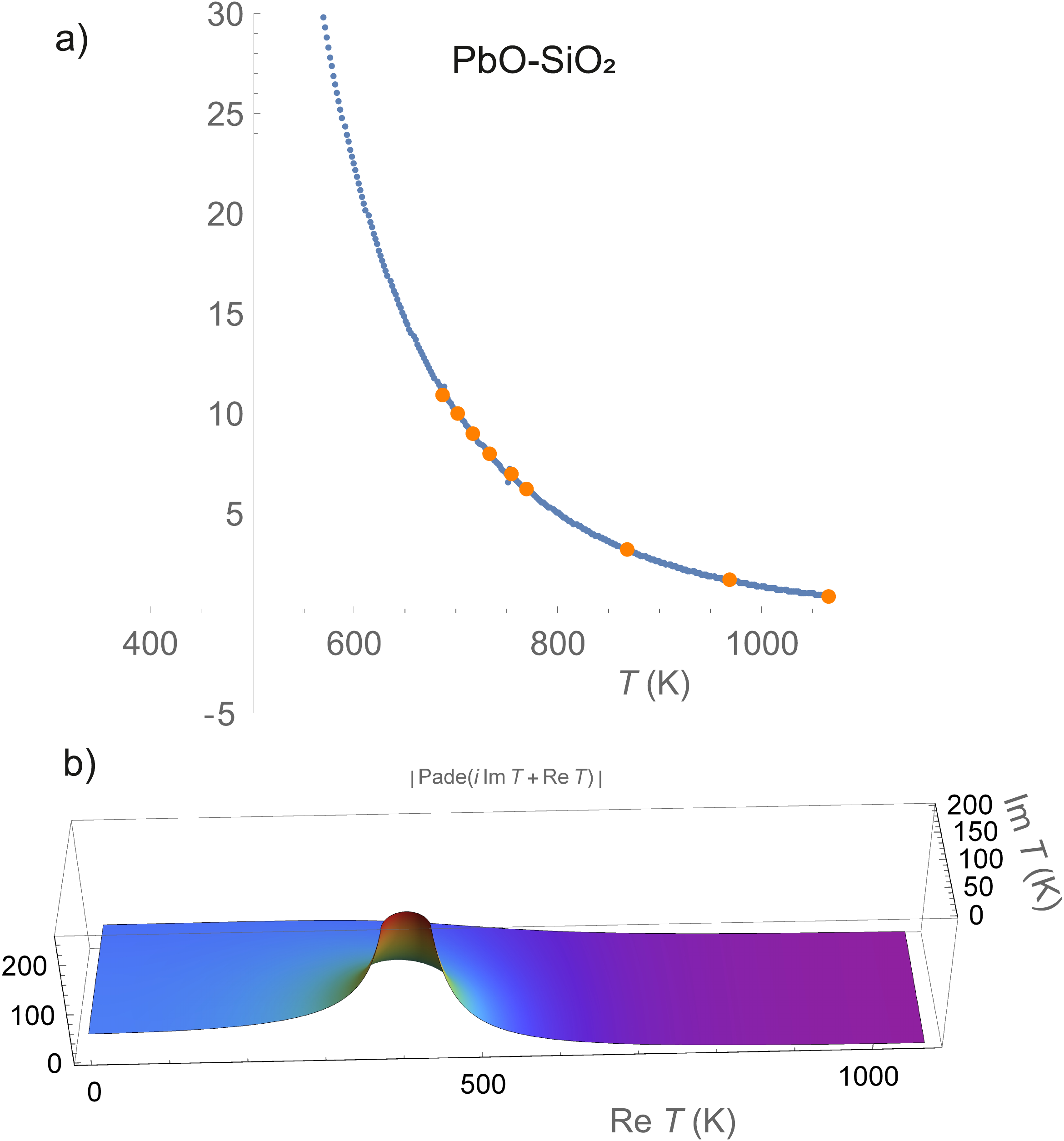}
  \caption{(Color online) a) Pade extrapolation of fragile $\mathrm{PbO-SiO}_2$ glass (blue curve). Orange dots show experimental data. b)  Pade approximant extrapolated into complex plane of temperature (temperature given in Kelvin units). The pole-singularity corresponds to $T_c$. }\label{figPbOSiO2}
\end{figure}

\section{Methods}

Pade approximants allow to interpolate and extrapolate a function specified by a table of $N$ points and, very importantly, to carry out an analytic continuation into the complex plane of the argument of the approximated function. The Pade approximant is defined through the rational function defined as the ratio of two polynomials. On the other hand, a rational function can be represented by a continued fraction, which approximates the function much more accurately compared to the series expansion. To construct the approximant, we use the ``multipoint algorithm''~\cite{Vidberg1977}. Suppose we know the values of a function $f (x_i) = u_i$ at the points $x_i$, where $ i = 1, 2, 3, ..., N$. Then the Pade approximant $P(x)$ is constructed as follows:
\begin{gather}
  P(x)=\frac{a_1}{\frac{a_2 (x-x_1)}{\frac{a_3 (x-x_2)}{\ldots}+1}+1},
\end{gather}
where $a_i$ are determined from the condition $P (x_i) = u_i$, which holds if $a_i$ satisfy the relations
\begin{gather}\label{eq5}
  a_i=g_i(x_i),\qquad g_1(x_i)=u_i,\qquad i=1,2,\ldots,N
  \\ \label{eq6}
  g_p(x)=\frac{g_{p-1}(x_{p-1})-g_{p-1}(x)}{(x-x_{p-1})g_{p-1}(x)},\qquad p\geq2.
\end{gather}
Expression \eqref{eq5} fulfills the role of the boundary condition for recursive relation~\eqref{eq6}. For example, taking $x = x_{i_0}$, then we find $g_1 (x_{i_0})$ from~\eqref{eq5} and $g_j (x_{i_0})$,
$j = 2, 3, ..., i$ from~\eqref{eq6}.

\section{Results}
\subsection{Repulsive shoulder system as the minimal model for glassformer}

First, we consider the one-component system of particles interacting through repulsive core-softened potential~\cite{Fomin2008JCP} (Repulsive Shoulder System (RSS)) as the model glassformer. Earlier, we have shown that RSS demonstrates excellent glass-forming ability in certain density domain~\cite{Ryltsev2013PRL}. The RSS potential has the following form:
\begin{gather}\label{csp}
  U(r)=\varepsilon\left(\frac{\sigma}{r}\right)^{n}+\varepsilon n_F\left[2k_0\left(r-\sigma_1 \right)\right].
\end{gather}
Here $n_F(x)=1/[1+\exp(x)]$.  The unit of energy --- $\varepsilon$, $\sigma$ and $\sigma_1$ are ``hard''-core and ``soft''-core scales. Taking these units, the dimensionless quantities are $\tilde{{\bf r}}\equiv {\bf r}/\sigma$, $\tilde U=U/\varepsilon$, temperature $\tilde T=T/\epsilon$, density $\tilde{\rho}\equiv N \sigma^{3}/V$, and time $\tilde t=t/[\sigma\sqrt{m/\varepsilon}]$, where $m$ and $V$ are the molecular mass and system volume. Later we omit the tildes. We also choose $n=14$, $k_0=10$, and $\sigma_1=1.35$.

The RSS system at given parameters shows glass formation even in the universal limit with the cooling rate $v\to 0$~\cite{Ryltsev2013PRL}. We took data from Ref.~\cite{Ryltsev2013PRL} on the diffusion coefficient $D(T)$  in the liquid phase not far above the glass transition threshold.  The chosen density $\rho=0.6$ corresponds to the region with high glass-forming ability at the RSS phase diagram. We approximated $D(T)$ using three different approaches: MCT, VF and Pade. After that low temperature extrapolation was performed to determine the glass transition temperature. The results are shown in Fig.~\ref{Pade_vs_MCT}. RSS is the superfragile glassformer~\cite{Angell1996JPC,Ryltsev2013PRL}, so VF approximation is quite accurate. That explains why Pade approximant in Fig.~\ref{Pade_vs_MCT} is close to VF. Fig.~\ref{Pade_vs_MCT} may be considered as the stress test of the suggested Pade extrapolation method.

Data for $D(T)$ extracted from  Ref.~\cite{Ryltsev2013PRL} has low accuracy. In Fig.~\ref{Pade_vs_MCT} this is even seen: the last data point corresponding to the lowest temperature is below the expected position. Such a low level of accuracy at first glance should prevent extrapolation by the analytical continuation method making it unstable. To solve this problem, we introduce averaging over two types of disorder into the calculation of the Pade approximant. We randomly remove 5\% of the data before constructing the Pade approximant. Then we average the result over a large number of random samples. Also, we add small Gaussian noise to the data and average the result over it. These procedures allows to build extrapolation in the presence of the inaccurate data in Fig.~\ref{Pade_vs_MCT}.

The MD-data~\cite{Ryltsev2013PRL} used to draw Fig.~\ref{Pade_vs_MCT2} was more accurate than used to prepare Fig.~\ref{Pade_vs_MCT}. We even did not have to average over the Gaussian noise and random data samples during the Padé approximant construction.

Using Pade extrapolation method we were able to extract the critical temperature $T_c$ where $D(T)$ goes to infinity. For fragile glassformers, this temperature is close to the so-called Kauzmann temperature~\cite{Kauzmann1948,Stillinger1988JCP,Angell1996JPC,Tanaka2003PRL,Kelton2016JPC}, $T_K$ [The Kauzmann temperature is defined by intersection of the crystal entropy curve with that extrapolated for the supercooled liquid~\cite{Stillinger1988JCP,Tanaka2003PRL}.] Usually $T_K\lesssim T_g$. In Figs.~\ref{Pade_vs_MCT}-\ref{Pade_vs_MCT2} we find $T_c$ as zero of Pade extrapolated $1/\log[D(T)]$. For RSS system we have got that $T_c$ calculated using VF and Pade nearly coincide. This is expected because RSS system is superfragile.

\section{Experimental glassformers}
\subsubsection{ Glycerol}

Now we take one of the most studied glass-forming systems, glycerol~\cite{Kauzmann1948,Bartos2001JPC}, to test our Pade-approximant method. According to Angell classification~\cite{Angell1996JPC},  glycerol is the fragile glassformer. So its viscosity varies with temperature  not according to the Arrhenius activation law (as in strong glassformers) but more or less according to VF or MCT.

We took experimental data for the viscosity of glycerol~\cite{Bartos2001JPC} and used them to build the Pade approximant, see Fig.~\ref{Pade_GL}(a), where the viscosity of glycerol is plotted in log scale. Orange spheres show the experimental data and the blue curve corresponds to Pade approximant. In Fig.~\ref{Pade_GL}(b) viscosity is extrapolated into the complex plane of temperature to detect $T_c$. It corresponds to the pole-singularity situated at the real axes of temperatures. We get $T_c\thickapprox 120$K. This value is close to $T_c$ found in~\cite{Bartos2001JPC}.

\subsubsection{ $\mathrm{SiO}_2$}
Now we consider $\mathrm{SiO}_2$ which is an representative strong glassformer according to Angell classification. Its viscosity processed by Pade-method is shown in Fig.~\ref{figSiO2}. Since the system demonstrates nearly Arrhenius temperature dependence of the viscosity, $T_c$ is not seen: Pade approximant is continuous at all temperatures.

\subsubsection{$\mathrm{ PbO-SiO}_2$}
Now we take the fragile glassformer $\mathrm{ PbO-SiO}_2$ whose $T_c=454$K has been estimated in Ref.~\cite{Ferreira2007JPCS} using  VF approximation.  We get  $T_c=420$K using Pade approximant, see Fig.~\ref{figPbOSiO2}. This result provides reasonable agreement.

\section*{Conclusions}

We propose a method for unambiguous extrapolation of liquid kinetic coefficients to freezing temperatures. Our approach is based on numerical analytical approximation and error correction algorithms that allow overcoming instabilities related to inaccuracy of input data. Using the proposed method one can determine the freezing temperature $T_g$ and the critical temperature $T_c$ where liquid kinetic coefficients go to zero.

\acknowledgments
We thank  R. Khusnutdinov for stimulating discussions. This work was supported by the Russian Science Foundation (grant 18-12-00438). The numerical calculations are carried out using computing resources of the federal collective usage center 'Complex for Simulation and Data Processing for Mega-science Facilities' at NRC 'Kurchatov Institute' (http://ckp.nrcki.ru/), supercomputers at Joint Supercomputer Center of Russian Academy of Sciences (http://www.jscc.ru) and 'Uran' supercomputer of IMM UB RAS (http://parallel.uran.ru).


%

\end{document}